\documentclass[a4paper,11pt]{article}
\pdfoutput=1 % if your are submitting a pdflatex (i.e. if you have
\usepackage{jheppub} % for details on the use of the package, please
\usepackage[T1]{fontenc} % if needed`
\usepackage{slashed}
\usepackage{booktabs}
\usepackage{mathrsfs}
\usepackage{floatrow}
\newfloatcommand{capbtabbox}{table}[][\FBwidth]

\title{\boldmath
The 750 GeV Diphoton excess, Dark Matter\\
and Constraints  from  the IceCube experiment 
}

\author[a]{Enrico Morgante,}
\author[a]{Davide Racco,}
\author[b]{Mohamed Rameez,}
\author[a]{and Antonio Riotto}

\affiliation[a]{
D\'epartement de Physique Th\'eorique and Center for Astroparticle Physics, \\ 
Universit\'e de Gen\`eve, 24 quai Ansermet, CH-1211 Gen\`eve 4, Switzerland}

\affiliation[b]{D\'epartement de Physique Nucl\'eaire et Corpusculaire, \\ 
Universit\'e de Gen\`eve, 24 quai Ansermet, CH-1211 Gen\`eve 4, Switzerland}

\emailAdd{enrico.morgante@unige.ch}
\emailAdd{davide.racco@unige.ch}
\emailAdd{mohamed.rameez@unige.ch}
\emailAdd{antonio.riotto@unige.ch}
\graphicspath{{./Figures/}}

\usepackage[dvipsnames]{xcolor}
\definecolor{EnricoColour}{rgb}{0.1,0.7,0.1}
\definecolor{DavideColour}{rgb}{0.65,0.,0.5}
\definecolor{RameezColour}{rgb}{0.1,0.2,0.9}
\definecolor{ToniColour}{rgb}{0.9,0.,0.}

\newcommand{\new}[1]{#1}

\renewcommand{\d}{\mathrm d}
\renewcommand{\deg}{^\text{o}}
\renewcommand{\S}{P} % in case we want to label it as P
\newcommand{\mS}{m_\S}
\newcommand{\GS}{\Gamma_\S}
\newcommand{\cgg}{c_{gg}}
\newcommand{\cBB}{c_{BB}}
\newcommand{\Ggg}{\Gamma_{gg}}
\newcommand{\Ggaga}{\Gamma_{\gamma\gamma}}
\newcommand{\GZZ}{\Gamma_{ZZ}}
\newcommand{\GZga}{\Gamma_{Z \gamma}}
\newcommand{\Guu}{\Gamma_{u\overline u}}
\newcommand{\Gbb}{\Gamma_{b\overline b}}
\newcommand{\Gtt}{\Gamma_{t\overline t}}

\newcommand{\yxS}{y_{\chi\S}}
\newcommand{\yqS}{y_{q\S}}
\newcommand{\yuS}{y_{u\S}}
\newcommand{\ydS}{y_{d\S}}
\newcommand{\ysS}{y_{s\S}}
\newcommand{\ybS}{y_{b\S}}
\newcommand{\ytS}{y_{t\S}}
\newcommand{\mx}{m_\chi}
\newcommand{\mf}{m_\phi}
\newcommand{\tW}{\theta_W}
\newcommand{\AMS}[1]{\big| \mathcal M_{\S \to #1} \big| ^2}
\newcommand{\simplify}[1]{\textcolor{gray!90}{#1}}
\newcommand{\cPPff}{c_{PP\phi\phi}}
\newcommand{\cHHff}{c_{HH\phi\phi}}
\newcommand{\Gcap}{\Gamma_\text{cap}}
\newcommand{\Gann}{\Gamma_\text{ann}}
\newcommand{\spx}{\sigma_{p\chi}}
\newcommand{\spf}{\sigma_{p \phi}}
\newcommand{\vDM}{v_\text{DM}}
\newcommand{\mDM}{m_\text{DM}}
\newcommand{\ODM}{\Omega_\text{DM}}
\newcommand{\asv}{\langle \sigma v\rangle}
\newcommand{\bes}[1]{\mathscr K_{#1}}
\newcommand{\vMol}{v_\text{M\o l}}
\newcommand{\svx}[1]{\sigma(\overline \chi \chi\to #1) \cdot \vMol}
\newcommand{\svf}[1]{\sigma(\overline \phi \phi\to #1) \cdot \vMol}
\newcommand{\Eg}{E_\gamma}

\abstract{
Recent LHC data show hints of a new resonance in the diphoton distribution at an invariant mass of 750 GeV. Interestingly, this new particle
might  be both CP odd and  play the role of a portal into the dark matter sector.  Under these assumptions and motivated by the fact that 
the  requirement of $SU(2)_L$ invariance automatically implies the coupling of this alleged new resonance to $ZZ$ and $Z\gamma$, 
we investigate the current and future constraints
coming from the indirect searches performed through the neutrino telescope IceCube, \new{supplementing  them with direct detection experiments and $\gamma$-ray observations.} 
We show that 
\new{IceCube} constraints can be stronger than the ones from direct detection experiments if the dark matter mass  is larger  than a few hundred GeV. Furthermore, in the scenario in which the dark matter is a scalar particle, the IceCube data limit
 the cross section between the DM and the proton to values close to the predicted ones for natural values of the parameters.
}

\begin{document} 
\maketitle
\flushbottom

\noindent

\section{Introduction}
\label{sec: introduction}
The ATLAS and CMS collaborations have recently announced \cite{Atlas:2015diphoton, CMS:2015dxe} (for a recent update see Refs.~\cite{Atlas:2016diphoton,CMS:2016diphoton}) the observation of an excess of events in the search for two photons in the final state. The shape of the excess suggests the detection of a resonance at an invariant mass of approximately $750$ GeV, and preliminary analyses suggest a rather large value for the width of the resonance, around 45 GeV (equivalent to 6\% of its mass). 

The statistical significance of this observation is still far from conclusive. 
The ATLAS collaboration, with $3.2\, \text{fb}^{-1}$ of data, claims a statistical significance of $3.9\sigma$ (or $2.3\sigma$ by taking into account the look-elsewhere effect) with an excess of about 14 events, corresponding to a cross section of about $10\pm 3$ fb. 
The CMS collaboration partially supports this observation, with a weaker statistical significance (local significance of $2.9\sigma$) due to a smaller integrated luminosity of $2.6\, \text{fb}^{-1}$, and with an estimated cross section of $6\pm 3$ fb.

%\davide{Are there more updated statistical analyses? I copied the values of \cite{Giudice2015}. We must be more precise in stating what assumptions give that significance (narrow width approximation).}\\
If the excess were confirmed by the data collected in the continuation of the Run-2 of the LHC experiments, this discovery would represent a historical cornerstone in the investigation of fundamental interactions beyond the Standard Model (SM). 
%The observed decay channel into two photons already constrains the spin of the resonance (which we denote by $\S$) by virtue of the Landau-Yang theorem \cite{Landau:1948kw,Yang:1950rg}, forcing its spin to be either 0 or 2. 

It is tempting, although speculative, to try to relate this hypothetical new particle to  the fact that about 30\% of the energy density of the universe seems to be in the form of Dark Matter (DM) particles.
If the indication on the total width $\GS$ of the resonance persists, then the possibility of a coupling of this new resonance to DM with a sizeable branching ratio (BR) would gain stronger support, because it would be easier to obtain such a large width in a simple, weakly coupled model.
In that case, a natural scenario would be that $\S$ acts as a portal between the SM and DM \cite{Giudice2015,Mambrini:2015wyu, Backovic:2015fnp, Knapen:2015dap, Han:2015cty, Bi:2015uqd, Ghorbani:2016jdq, Bhattacharya:2016lyg, D'Eramo:2016mgv, Han:2015dlp, Dev:2015isx, Han:2015yjk, Park:2015ysf, Berlin:2016hqw, Borah:2016uoi, Ko:2016wce, Yu:2016lof, Okada:2016rav, Ge:2016xcq, Redi:2016kip}. 
Given that the  mass scale of the resonance is above the electroweak scale, it is reasonable to write an effective theory describing the interaction between $\S$ and the photon in an $SU(2)_L$ invariant way. In this case, an interaction of $\S$ with photons automatically implies the existence of the vertices $\S ZZ$ and $\S Z\gamma$, with fixed couplings (see Ref. \cite{ATLAS:Zgamma} for a Run-2 ATLAS analysis analysis of $Z\gamma$ final states). This very general statement has important implications for phenomenology, first of all at LHC but also in DM searches, if indeed $\S$ acts as a portal into the  DM sector.

The observed decay channel into two photons already constrains the spin of the resonance (which we denote by $\S$) by virtue of the Landau-Yang theorem \cite{Landau:1948kw,Yang:1950rg}, forcing its spin to be either 0 or 2.\footnote{Consequences of the Landau-Yang theorem can be evaded in scenarios in which a vectorial resonance decays via a cascade into a final state of three photons, two of which are too collimated to be discriminated in the detector (see e.g. Ref.~\cite{Chala:2015cev}). 
For a discussion about the fate of Landau-Yang theorem in non-Abelian gauge theories, see \cite{Cacciari:2015ela}.} 
In this work, we focus on the possibility that $\S$ is a particle of spin zero, which interacts with a DM particle charged under a $\mathbb Z_2$ symmetry that prevents it from decaying and makes it a viable DM candidate. 
We consider the two cases of a Dirac fermion $\chi$ and of a complex scalar $\phi$. 
The discussion of the DM phenomenology of this model is strongly affected by the assumption on the CP properties of $\S$. 
At the energy scales involved at the LHC, the production cross section of $\S$ is basically independent of whether $\S$ is a scalar or a pseudoscalar particle, whereas at low energies the two options bring to different non-relativistic effective operators. 

In the scalar case, the effective operator for the interaction between DM and nuclei is spin independent (SI), yielding strong bounds from direct detection (DD) experiments.
If however $\S$ is a pseudoscalar particle, then the low energy effective interaction between DM and a nucleus is spin dependent (SD).
%is given by $(\vec s_\chi \cdot \vec q)(\vec s_N \cdot \vec q)$, where $\vec s_\chi,\, \vec s_N$ are respectively the spins of $\chi$ and of the nucleus, and $\vec q$ is the momentum exchange. 
%Thus the interaction between DM and nuclei is both spin and velocity suppressed. 
In this case, the exclusion reach of DD is much weaker and has to be complemented with the constraints coming from indirect detection (ID), because the DM annihilation at low velocities occurs through an $s$-wave for the pseudoscalar case (while in the scalar case it occurs through a $p$-wave process).

In this paper we assume that the particle $\S$ is a pseudoscalar\footnote{See  Refs. \cite{Brustein:1999it, Ben-Dayan:2016gxw, Pilaftsis:2015ycr} for a discussion of a pseudoscalar field coupled to $B \widetilde B$ as a heavy hypercharge axion.} and  assess the  constraints coming from the IceCube (IC) experiment \cite{IceCubeSearchMethodPaper}, a neutrino telescope that can be used to study the DM annihilations occurring in the Sun \cite{ICRCSolarDM, Montaruli:2015six}.
DM particles can get captured in the gravitational well of the Sun if they scatter with atomic nuclei inside it and they lose some energy. 
The accumulation of DM particles is partly compensated by their annihilation, until the DM density in the Sun reaches an equilibrium level. 
If this is already achieved today, then one  can directly relate the annihilation rate to the capture rate in the Sun and constrain the interactions of the DM with the 750 GeV resonance as well as the interactions of the resonance with the SM particles.

%(details are provided in sec.~\ref{sec: icecube}). 
%The annihilation products of DM pairs always include neutrinos after secondary decays, even more so when we take into account the electroweak corrections, which are important at energies above the weak scale. \\
The following features  make particularly  interesting the  study of the  750 GeV resonance with the IC experiment, if the pseudoscalar
 $\S$ acts like a portal to DM:

\begin{itemize}
\vspace{-0.3em}
\item The decay of $\S$ into $ZZ$ and $Z\gamma$, granted by the assumption of $SU(2)_L$ invariance, ensures the presence of at least two annihilation channels for DM that yield energetic neutrinos which can be observed by IC.
\vspace{-0.3em}
\item Given the assumption of equilibrium inside the Sun, neutrino fluxes at Earth depend only on the BR's of DM annihilating into SM primary products. 
In the expression of the BR's, interesting simplifications occur and the bound from IC turns out to be independent of the total width $\GS$, of the coupling of $\S$ to the DM particle, and of the mass $\mS$ of the resonance.
Furthermore, the expressions of BR's can be rewritten as functions of the partial decay widths of $\S$, which are the quantities that can be directly measured via a resonant production at LHC.
\vspace{-0.3em}
\item IC constraints reach the highest exclusion power for DM masses of order $10^{-1}$ to  $1$ TeV \cite{ICRCSolarDM}. Indeed, for  DM masses higher than a TeV, the  DM number density in the Sun is reduced and this, together with the fact that primary neutrinos from DM annihilations hardly escape without interacting within the Sun and losing much of their energy, deteriorates the limit. At lower DM masses, the angular resolution of IC is poorer because of the lower number of Cherenkov photons produced by muons at these energies.\\
Interestingly, this mass window coincides with the order of magnitude $\mS$. Thus, if the DM particle has a mass close to $\mS$, IC might  give the strongest bounds on DM.
\vspace{-0.3em}
\end{itemize}
This paper is structured as follows. 
Sec.~\ref{sec: model} describes the different models we consider, and the benchmarks we choose for the couplings of $\S$ to SM particles which are consistent with LHC observations. 
Sec.~\ref{sec: icecube} illustrates the physics that links the observations of IceCube and DM annihilations, and the importance of electroweak (EW) corrections at the energy scales of interest to IC. 
Sec.~\ref{sec: results} contains the results for the constraints from IceCube, DD \new{and $\gamma$-rays observations (Fermi-LAT, HESS)} on the benchmark models we consider. Finally, in  Sec.~\ref{sec: conclusions} we summarise our results.

\section{The 750 GeV resonance as a portal to DM: the models}
\label{sec: model}
% S is pseudoscalar real field, with transformation properties $\S\rightarrow -\S$ under $P$ and $\S\rightarrow \S$ under $C$
The observation of $\S$ in final states with two photons forces $\S$ to be coupled at least to the photon field strength. According to our assumption,  $\S$ is a pseudoscalar and the effective vertex for its interactions with photons  must be of the form $\S F_{\mu\nu}\widetilde F_{\mu\nu}$ in order not to introduce a source of CP violation, where $\widetilde F_{\mu\nu}=\frac 12 \varepsilon_{\mu\nu\rho\sigma} F^{\rho\sigma}$ is the dual field strength. 

Since the theory must have a cut-off at least higher than $\mS=750$ GeV,  it is more than reasonable to write down an explicitly $SU(2)_L$-invariant Lagrangian. 
Given the current lack of excess in $WW$, $ZZ$ and $Z\gamma$ searches, the coupling of $\S$ to $B\widetilde B$ is favoured with respect to a coupling to $W^i\widetilde W^i$ (where $i=1,2,3$ is the $SU(2)_L$ index). 
Thus, for minimality, we set to zero the coefficient of the vertex $\S W^i\widetilde W^i$.\footnote{This can be easily realised in a UV completion where the effective vertex between $\S$ and the vector bosons arises via a fermion loop with fermions charged under $U(1)_Y$ and singlets under $SU(2)_L$.}

The absence of anomalies in the diphoton searches of Run-1 at LHC strongly favours the hypothesis that $\S$ couples also to gluons\footnote{See also  Ref. \cite{Chu:2012qy} for a discussion about a DM candidate coupling to both photons and gluons via loop interactions.}
 and/or quarks, because the parton distribution function (PDF) for photons changes only mildly from 8 TeV to 13 TeV, whereas the experimental results of Run 1 and 2 can be consistent if $\S$ is produced mainly by gluon fusion or from heavy quarks (see for instance Ref.~\cite{Giudice2015}).
We ignore the coupling of $\S$ to leptons, since this is irrelevant for diphoton production and has a minor impact for IC and DD.
We consider therefore the following Lagrangian density
\begin{equation}
\begin{aligned}
\mathcal L = & \mathcal L_\text{SM} +\frac 12 (\partial_\mu \S)(\partial^\mu \S)- \frac 12 \mS^2 \S^2
   + \frac{\cgg}{\Lambda}\, \S \, G^a_{\mu\nu} \widetilde G^{a\, \mu\nu}
   + \frac{\cBB}{\Lambda}\, \S \, B_{\mu\nu} \widetilde B^{\mu\nu} \\
      & + i\, \S \sum_{i=1}^3\left(  \frac{y_{d^i\S} H}{\Lambda} \overline Q^i_{L} d^i_{R}  +
      \frac{y_{u^i\S} H^\text c}{\Lambda} \overline Q^i_{L} u^i_{R}  
      \right) \new{+\text{ h.c.} }
\\
      & + \mathcal L_{\rm DM}  \,,
\end{aligned}
\end{equation}
%      & + i\, \S \left( \sum_q \frac{\yqS H}{\Lambda} \overline q_L q_R  +\text{h.c.} \right) + i\, \S \left( \sum_q \frac{\yqS \, i\sigma_2 H}{\Lambda} \overline q_L q_R  +\text{h.c.} \right) \\
where $\Lambda$ is the dimensionful scale of the effective theory, $a=1,\dots,8$ is the gluon $SU(3)_c$ index, $i$ is the family index, $H$ is the Higgs doublet and $H^\text c=i\sigma_2 H^*$ is its conjugate, $Q_{i\, L}$ is the quark weak doublet, and $\cgg,\, \cBB,\, \yqS, \, \yxS$ are real coefficients.

As for the DM Lagrangian we envisage two possible (and mutually excluding) cases: either the DM   is a Dirac fermion $\chi$  (the results being analogous in case of a self-conjugate DM particle)
\begin{equation}
\label{eq:L DM fermion}
\mathcal L_{\rm DM}= \overline \chi (i \slashed \partial -\mx) \chi + \yxS \, \S \, \overline \chi i \gamma^5 \chi \,,
\end{equation}
or it is a scalar particle $\phi$ with CP-conserving Lagrangian
\begin{equation}
\label{eq:L DM scalar}
\mathcal L_{\rm DM}=  (\partial_\mu \phi^*)(\partial^\mu \phi) -m_\phi^2 \, |\phi|^2 +A_P \S \,|\phi|^2
\new{+ \cPPff \S^2 |\phi|^2  + \cHHff |H|^2 |\phi|^2} \,,
\end{equation}
where $A_P$ has to be thought of as a spurion field which changes sign upon CP (for example it might be the vacuum expectation value of some heavy parity-odd field). 

Notice that in both cases the DM  could be identified with one of those particles in the fermion or scalar  multiplets coupled to the resonance and which, upon integrating them out,  give rise to the effective interaction between $P$ and   photons and gluons. If so, the DM  must be of course a  singlet under the SM gauge group.

\new{
The couplings $\cPPff$ and $\cHHff$ of Eq.~\eqref{eq:L DM scalar} are not currently constrained  by the observation of the decay of $P$ into two photons, thus they cannot be linked to the quantities measured by the experimental collaborations. 
Furthermore, the term $P^2 |\phi|^2$ does not alter quantitatively the branching ratios of the annihilation of $\overline \phi \phi$ to SM particles relevant for the bounds derived from IceCube (see also footnote~\ref{footnote:DM DM to P P}). 
Since we want to exploit the information collected by the experimental collaborations on the allowed partial decay widths of $\S$, and we do not want to introduce too many free parameters in the benchmark choices we are going to illustrate,  we decided to set $\cPPff=\cHHff=0$\footnote{If these parameters are not vanishing, one should correspondingly include them in the computation of the relic abundance.}.
}

In order to explain the diphoton excess and ensure that it can be consistent with Run-1, we need at least $\cBB$ and one among $\cgg$ and $\yqS$ to be different from zero. 
In the benchmark scenarios that we illustrate below we always assume a coupling of $\S$ both to gluons and to light quarks. 
These allow respectively to improve the compatibility between Run-1 and Run-2 of LHC (given the stronger increase of gluon PDF with energy with respect to valence quarks), and to increase the elastic cross section between proton and DM, in order to achieve more easily the equilibrium of DM number density in the Sun.

The coupling of $\S$ to $B\widetilde B$ induces partial decay widths of $\S$ to the final states $\gamma\gamma$, $Z\gamma$ and $ZZ$, proportional respectively to $\cos^4\theta_W$, $4\cos^2\theta_W \sin^2\theta_W$, $\sin^4\theta_W$, where $\theta_W$ is the Weinberg angle. We stress again that the compulsory coupling of $\S$ to $\gamma \gamma$ automatically implies, when requiring $SU(2)_L$ invariance, a coupling of $\S$ to $ZZ$ and $Z\gamma$ that can give a relevant signal for IC, given the typical hard spectrum of $\nu$ from $Z$ decay.

We identify then three benchmark scenarios, where we always assume a coupling of $\S$ to DM, gluons, photons (and hence $ZZ$ and $Z\gamma$), a light quark (for simplicity we assume that only $\yuS\neq 0$, but the result would not differ much if also $\ydS$ and $\ysS$ were present), and possibly a heavy quark (top or bottom).

We choose the couplings $\cBB,\,\cgg,\, \yuS,\, \ybS,\, \ytS,$ and the coupling to DM in such a way to satisfy three requirements: \textbf{1)} the corresponding partial decay widths of $\S$ satisfy the present collider bounds from Run-1 and explain the Run-2 diphoton excess \cite{Giudice2015}, \textbf{2)}  the capture rate of DM in the Sun due to elastic scattering with the proton is high enough so that the DM number density in the Sun has reached equilibrium today, and \textbf{3)} we obtain a reasonable compatibility between $\gamma \gamma$ searches in Run-1 and Run-2\footnote{We quantify this latter requirement by means of the gain factors defined as the ratio $r_\wp$ between the cross section for the process $p p \to \S \to \gamma\gamma$ at 13 and 8 TeV, computed assuming that the resonance is produced from a single parton $\wp$. Compatibility between Run-1 and Run-2 is achieved for $r_\wp\sim 5$, which favours $\wp=g,b,s,c$ and disfavours $\wp=u,d,\gamma$.}.

The preliminary indication on the total decay width $\GS=0.06\, \mS$ is included in most of the bounds shown in Ref. \cite{Giudice2015}.  
As we will see in section ~\ref{sec: icecube}, the  IC bounds depend only on the BR's of $\overline \chi \chi$ (or $\overline \phi \phi$) into SM particles. With the use of Eqs.~\eqref{eq:width P to ij} and \eqref{eq:xsec chi chi to i j} (or \eqref{eq:xsec phi phi to i j} in the scalar DM case) one can rewrite the BR into, say, $\gamma\gamma$, as follows, if $\S$ couples for example to $BB$, $gg$ and $\overline u u$:
\begin{equation}
\label{eq:BR scenario A}
\text{BR}\,(\text{DM DM} \to \gamma\gamma) = \left( 
  1 + \frac{\GZga}{\Ggaga}  + \frac{\GZZ}{\Ggaga}  + \frac{\Ggg}{\Ggaga}  + \frac{\Gamma_{\overline uu}}{\Ggaga}
  \right)^{-1} \,.
\end{equation}
As stressed in appendix~\ref{sec:appendix A}, the dependence on $\GS$ and on the coupling of $\S$ to DM disappears rendering  the IC flux predictions independent from both. 

Even more importantly, the ratios of partial widths appearing in Eq.~\eqref{eq:BR scenario A} are the quantity directly constrained by LHC observations. 
Furthermore, the allowed ranges for these ratios do not alter significantly if one drops out the assumption $\GS=0.06\, \mS$ (see for example Figs.~1 and 2 of Ref.~\cite{Giudice2015}). 

We now illustrate the three benchmark scenarios we have chosen\footnote{For the rest of this section, for simplicity of notation we denote the DM particle by $\chi$, but the discussion is identical for the case of scalar DM.}.
\begin{itemize}
\begin{figure}[h!]
\vspace{-1em}
\begin{floatrow}
\capbtabbox[0.35\textwidth]{
  \begin{tabular}{lc}
  \toprule
  $\Ggaga \,/\,\mS$ & $2.7\cdot 10^{-4} $\\
  $\Ggg\,/\,\mS$ & $2.2\cdot 10^{-4} $\\
  $\Guu\,/\,\mS$ & $1.1\cdot 10^{-4} $\\
  $\Gbb\,/\,\mS$ & $0 $\\
  $\Gtt\,/\,\mS$ & $0 $\\
  \bottomrule
  \end{tabular}
}{
  \caption{Partial decay widths of $\S$ into SM channels for scenario A.}
  \label{tab:scenario A}
}
\ffigbox[0.6\textwidth]{
    \includegraphics[width=0.6\textwidth]{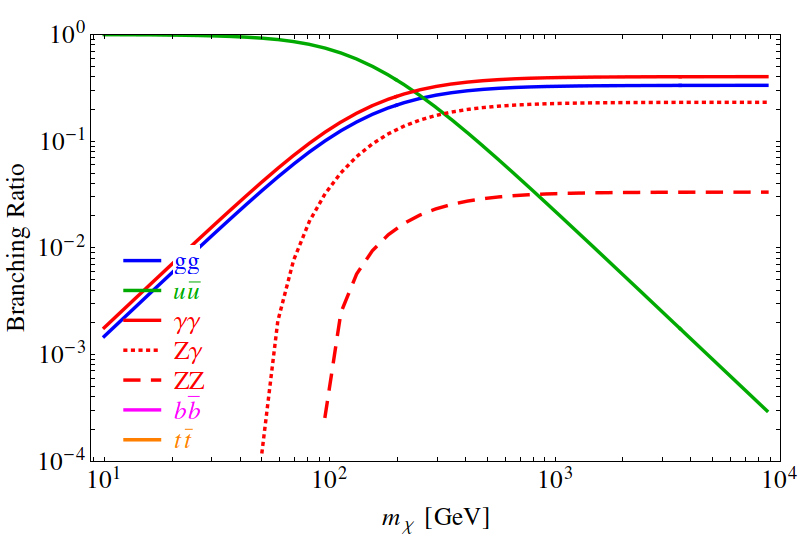}
    \vspace{-1.5em}
}{
    \caption{Branching ratios for the annihilation of $\overline \chi\chi$ into SM channels as a function of $\mx$ for scenario A.}
    \label{fig:BR A}
 }
\end{floatrow}
%\end{figure}
%%
%\begin{figure}[h!]
%\vspace{-1em}
\begin{floatrow}
\capbtabbox[0.35\textwidth]{
  \begin{tabular}{lc}
  \toprule
  $\Ggaga \,/\,\mS$ & $6.1\cdot 10^{-4} $\\
  $\Ggg\,/\,\mS$ & $2.3\cdot 10^{-5} $\\
  $\Guu\,/\,\mS$ & $1.2\cdot 10^{-5} $\\
  $\Gbb\,/\,\mS$ & $4.8\cdot 10^{-3} $\\
  $\Gtt\,/\,\mS$ & $0 $\\
  \bottomrule
  \end{tabular}
}{
  \caption{Partial decay widths of $\S$ into SM channels for scenario B.}
  \label{tab:scenario B}
}
\ffigbox[0.6\textwidth]{
    \includegraphics[width=0.6\textwidth]{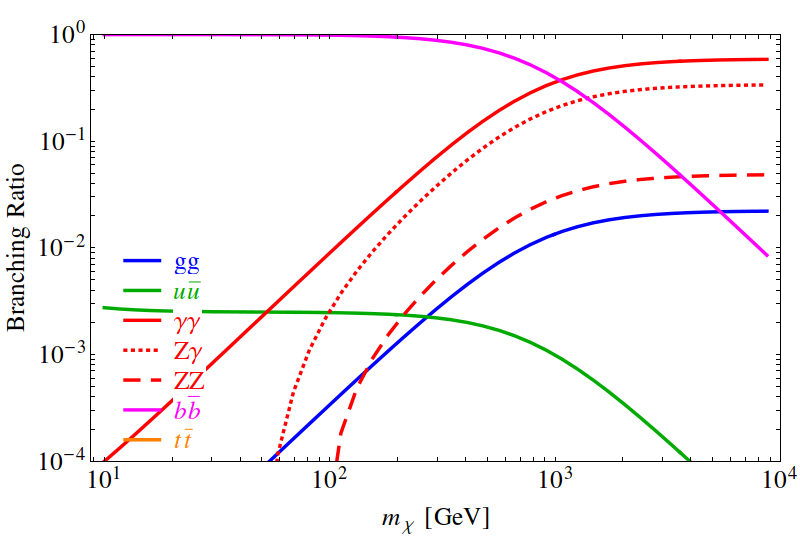}
    \vspace{-1.5em}
}{
    \caption{Branching ratios for the annihilation of $\overline \chi\chi$ into SM channels as a function of $\mx$ for scenario B.}
    \label{fig:BR B}
 }
\end{floatrow}
%\end{figure}
%%
%\begin{figure}[h!]
%\vspace{-1em}
\begin{floatrow}
\capbtabbox[0.35\textwidth]{
  \begin{tabular}{lc}
  \toprule
  $\Ggaga \,/\,\mS$ & $2.7\cdot 10^{-4} $\\
  $\Ggg\,/\,\mS$ & $2.2\cdot 10^{-4} $\\
  $\Guu\,/\,\mS$ & $5.8\cdot 10^{-6} $\\
  $\Gbb\,/\,\mS$ & $0 $\\
  $\Gtt\,/\,\mS$ & $2.6\cdot 10^{-2}$\\
  \bottomrule
  \end{tabular}
}{
  \caption{Partial decay widths of $\S$ into SM channels for scenario C.}
  \label{tab:scenario C}
}
\ffigbox[0.6\textwidth]{
    \includegraphics[width=0.6\textwidth]{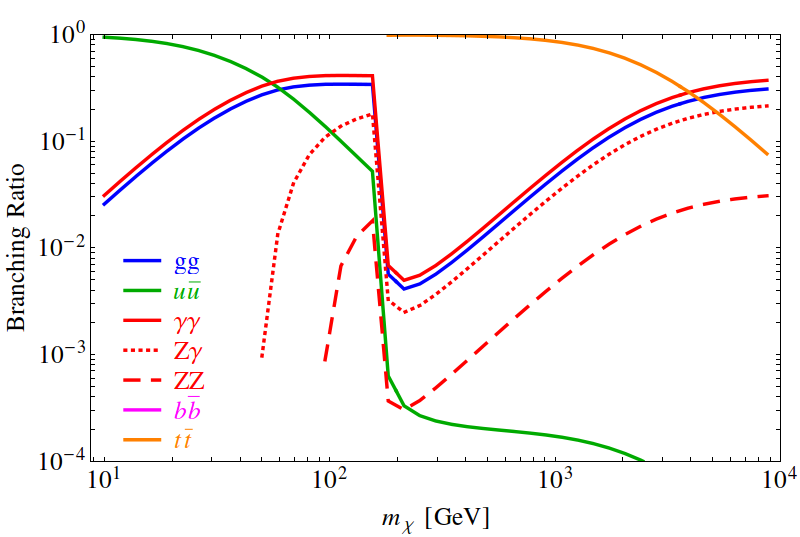}
    \vspace{-1.5em}
}{
    \caption{Branching ratios for the annihilation of $\overline \chi\chi$ into SM channels as a function of $\mx$ for scenario C.}
    \label{fig:BR C}
 }
\end{floatrow}
\end{figure}
  \item[$\diamond$] \textbf{Scenario A}: $\S$ couples to $B$, $g$, $u$, $\chi$. We choose the coefficients of the model leading to the partial decay widths of $\S$ into SM particles listed in table~\ref{tab:scenario A}. 
  The production of $\S$ at LHC at 13 TeV occurs mainly from gluons, while 
  at low energies the elastic scattering with protons is mediated mainly by $u$. The resulting BR's  for the annihilation of $\overline \chi \chi$ into SM channels as a function of $\mx$ are shown in Fig.~\ref{fig:BR A}.\footnote{\label{footnote:DM DM to P P}
  We do not show the BR for the process $\overline\chi \chi \to \S\S $, where $\chi$ is exchanged in the $t$-channel, 
  %, for a twofold reason: from a technical point of view, the software PPPC4DM ID \cite{Baratella:2013fya} does not allow to consider DM annihilations into final states with more than two particles. More significantly, 
  because the result in terms of SM final particles stays identical: the decay products of $\S\S$ have the same identical BR's as the ones of $\overline \chi \chi$ into two-body SM final states.}
  \newline
  The relevant channels for the production of hard neutrinos are $ZZ$ and $Z\gamma$.
  \item[$\diamond$] \textbf{Scenario B}: $\S$ couples to $B$, $g$, $u$, $\chi$ and $b$. The chosen partial decay widths of $\S$ are listed in table~\ref{tab:scenario B}.
  The coupling of $\S$ to $b$ quarks improves the compatibility between Run-1 and Run-2, although the coupling to photons is quite larger than the one to gluons. As far as LHC is concerned, the production of $\S$ occurs mainly from $b$ partons.
  The resulting BR's for the annihilation of $\overline \chi \chi$ into SM channels are shown in Fig.~\ref{fig:BR B}.
  \newline
  The $\overline bb$ channel yields soft neutrinos for IceCube, for which the main channels remain $ZZ$ and $Z\gamma$.
  \item[$\diamond$] \textbf{Scenario C}: $\S$ couples to $B$, $g$, $u$, $\chi$ and $t$. The chosen partial decay widths of $\S$ are listed in table~\ref{tab:scenario C}.
  We choose the coupling to $\overline tt$ to be the main one. This channel does not contribute to the production at LHC, and amplifies the signal for IC. 
  %The coupling to gluons is then required to be not too small in order not to spoil compatibility between Run-1 and Run-2.
  The production at LHC occurs via gluon fusion.
  The corresponding BR for $\overline \chi \chi$ into SM channels are plotted in Fig.~\ref{fig:BR C}.\\
  Given its large BR, $\overline tt$ is the only important channel for IceCube.
\end{itemize}
We illustrate in the next section the main features of IceCube and its relevance in DM searches. Section~\ref{sec: results} shows the resulting bounds from IC, DD experiments \new{and $\gamma$-rays observations} for the three scenarios just described.

\section{Constraints on DM annihilations from the IceCube experiment}
\label{sec: icecube}

The IceCube experiment, located at the South Pole, is a neutrino telescope observing high energy neutrinos by detecting Cherenkov photons radiated by charged particles produced in their interactions \cite{Achterberg:2006md}. Muons from $\nu_\mu$ and $\overline \nu _\mu$ charged current interactions leave long visible tracks within the detector, which can be easily reconstructed to estimate the direction of the incoming neutrino. IceCube has an angular resolution of a few degrees for $\sim$ 100 GeV $\nu_\mu$ (and $< 2^{\circ}$ for $\sim$ for a 1 TeV $\nu_\mu$), allowing it to search for an excess of GeV-TeV neutrinos from the direction of the Sun \cite{Aartsen:2012kia}.

DM accumulates in celestial bodies if it loses some of its kinetic energy via elastic scattering and remains gravitationally bound within them. As the DM number density $n_{\rm DM}$ increases, pairs of DM particles annihilate into SM ones, and some are lost due to evaporation  (for further details, see Ref.~\cite{Jungman:1995df}  and references  therein). The system   eventually reaches an equilibrium, and  $n_{\rm DM}$ freezes.
 Neglecting the evaporation rate, which is a plausible assumption for $m_\chi \gtrsim 10$ GeV, at equilibrium the capture rate $\Gcap$ is equal to $2 \Gann$, where $\Gann$ is the annihilation rate of DM particles into SM particles and is proportional to $n_{\rm DM}^2$. 
 The capture rate $\Gcap$ is proportional to the elastic scattering cross section $\spx$ between a proton and the DM particle $\chi$. 
The higher $\Gcap$, the faster the equilibrium is reached.
When equilibrium is achieved, $\spx$ can be directly related to the annihilation rate. 

The key point to note is that once equilibrium is attained, in order to predict the neutrino fluxes searched for by IC  only  the ratios of the annihilation cross sections (or branching ratios) matter, and these ratios can be directly related to LHC measured quantities.  Indeed, once we know the flux of neutrinos on Earth per DM annihilation, we can infer an upper bound on the rate of DM annihilation from the non-observation of an excess over the expected background. In other words, the neutrino fluxes depend only  on branching ratios, not on the absolute value of annihilation rate. 

The computation of the energy spectra  of the neutrinos  originated from the annihilation  of DM particles is performed including  the  electroweak corrections, which can  significantly alter the  neutrino spectra when the mass of the DM  particles is larger than the electroweak scale \cite{Ciafaloni:2010ti}. This is because soft electroweak gauge bosons are copiously radiated when the mass of the DM is larger than the 
gauge boson mass, and this opens new channels in the final states, including neutrinos,  which otherwise would be forbidden if such corrections are neglected. In order to implement such electroweak corrections, we make use of the PPPC4DM ID code \cite{Baratella:2013fya}.

The current most stringent bounds on spin dependent DM-nucleon scattering cross section from IC comes from a search employing 3 years of IC  data \cite{Montaruli:2015six}. The IC collaboration presents constraints in terms of three benchmark cases: DM annihilation into 
$b\bar{b}$, $W^+W^-$ and  $\tau^+\tau^-$. The least stringent constraints are for DM annihilating 100\% into $b\bar{b}$, a situation which produces few neutrinos and  at energies much below the mass of the DM. DM particles annihilating 100\% to $W^+W^-$ or $\tau^+\tau^-$ produce a significantly larger number of neutrinos at energies close to the DM mass and consequently stronger bounds on the scattering cross section. Neutrino flux predictions for these cases can be obtained from WimpSim \cite{Blennow:2007tw}.
\begin{figure}[h!]
\centering
\includegraphics[width=0.75\textwidth]{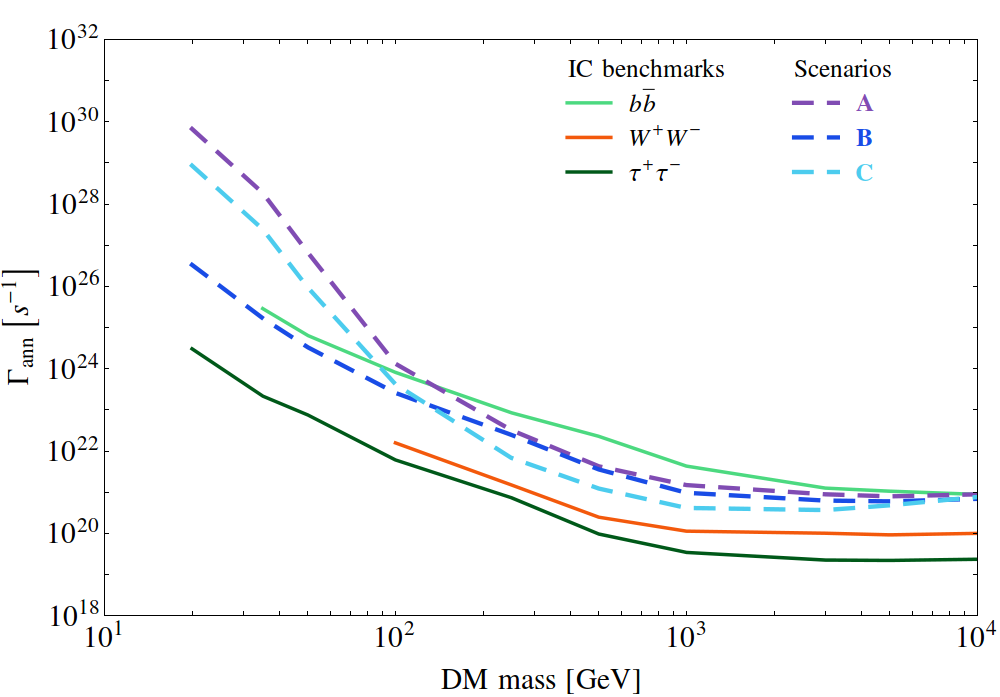}
\caption{Upper limits on the DM annihilation rate in the Sun for the IceCube benchmark channels, as well as the three scenarios discussed above. The benchmark limits shown and used for rescaling conservatively correspond to the upper edge of the systematics band in \cite{Montaruli:2015six} and are derived using the online flux conversion tool provided by the authors of WimpSim \cite{DMSimTool}.}
\label{fig:resultsIceCube}
\end{figure}
\newline
The event rate expected from a differential (anti)-muon neutrino flux $\mathcal{F}(E)$ in an IC sample of effective area $A_{\rm eff}$ is given by
\begin{equation}
n_{\rm s}\left(\mathcal{F}\right) = \int_{E_{\rm threshold}}^{\mx} \mathcal{F}(E) \cdot A_{\rm eff}(E) \, \d E,
\label{eq:nsintegral}
\end{equation}
while the median energy $E_{\rm med}(\mathcal{F})$ is defined such that
\begin{equation}
\int_{E_{\rm threshold}}^{E_{\rm med}} \mathcal{F} \cdot A_{\rm eff} \, \d E  = \int_{E_{\rm med}}^{\mx}  \mathcal{F} \cdot A_{\rm eff} \, \d E.
\label{eq:E_med}
\end{equation}
These quantities can be calculated for neutrino flux predictions calculated as described before, and also for the benchmark channel predictions, for each of the three IC samples described in Ref. \cite{ICRCSolarDM}. Subsequently, the limit on the annihilation rate $\Gamma_{\rm ann}$ for theoretical neutrino flux $\mathcal{F}(E)$ can be obtained by rescaling the benchmark limits according to the expression:
\begin{equation}
 \Gamma_{\rm ann}^{\rm theory} = \Gamma_{\rm ann}^{\rm benchmark} \cdot \frac{n_{\rm s}(\mathcal{F}_{\rm benchmark})}{n_{\rm s}(\mathcal{F}_{\rm theory})} \cdot \frac{\Psi\left(E_{\rm med}(\mathcal{F}_{\rm theory})\right)}{\Psi\left(E_{\rm med}(\mathcal{F}_{\rm benchmark})\right)},
\label{eq:scaling2}
\end{equation}
where $\Psi(E)$ is the angular resolution of the IceCube sample at at neutrino energy $E$. This scaling is possible because the search is performed using the unbinned maximum likelihood ratio method \cite{ICRCSolarDM, IceCubeSearchMethodPaper}, for which the sensitivity scales with ${\rm Signal}/\sqrt{\rm Background}$ and the background level varies as $\Psi^2$. In practice a scaling factor averaged over the three samples weighted by their exposure at the median energy is used. For a given $\mathcal F_{\rm theory}$, $\sigma_{\rm theory}$ can be calculated with respect to\ any of the three benchmark IC channels. The different calculations are consistent to within $\sim 30\%$ and are thus averaged.

The limits on $\Gamma_{\rm ann}$ for the IC benchmark channels can be obtained from the limits on $\sigma$ using tools provided by WimpSim and DarkSuSy \cite{DMSimTool}. Figure \ref{fig:resultsIceCube} illustrates the limits on $\Gamma_{\rm ann}$ for the scenarios discussed above, as well as for the IC benchmark cases.
\new{
We notice that the exclusion bounds reported in Fig.~\ref{fig:resultsIceCube} are not affected by the thresholds for the opening of new annihilation channels, such as $\mx\sim m_\text{top}$ in scenario C for example.
The reason is that the assumption of equilibrium in the DM number density in the Sun implies that IC constrains the DM capture rate in the Sun, rather than its annihilation rate.
This is the reason why also the corresponding bounds in Fig.~\ref{fig:results fermion DM} and \ref{fig:results scalar DM} do not display bumps in correspondence of kinematic thresholds.
}

\section{Results and discussions}
\label{sec: results}
Having set our benchmark models describing the interactions of the pseudoscalar $P$ with the SM particles and having addressed  the way we deal with the IC physics, we now proceed to present our results in the cases in which the DM is a fermion and a scalar.

\subsection{The case of fermionic DM}
Using the solar capture rates evaluated in Ref.~\cite{Catena:2015uha},  the bounds on $\Gamma_{\rm ann}$ shown in Fig.~\ref{fig:resultsIceCube} can be interpreted as a bound on $\spx$ for the operator $\mathcal O_6^\text{NR}=(\vec s_\chi \cdot \vec q)(\vec s_N \cdot \vec q)$. This operator   originates in the non-relativistic limit in the case in which the DM is a Dirac fermion $\chi$. 

In comparison to direct detection, bounds from solar DM searches are generally more stringent in the scenario where the DM-nucleon scattering depends on the spin of the nucleus, since the Sun consists of mostly protons, in contrast to the target nuclei used in direct detection experiments which usually have no spin. Bounds on $\spx$ for the operator $(\vec s_\chi \cdot \vec q)(\vec s_N \cdot \vec q)$ are significantly weaker than for the common spin dependent operator due to the double velocity suppression.

\begin{figure}[h!]
  \includegraphics[width=0.49\textwidth]{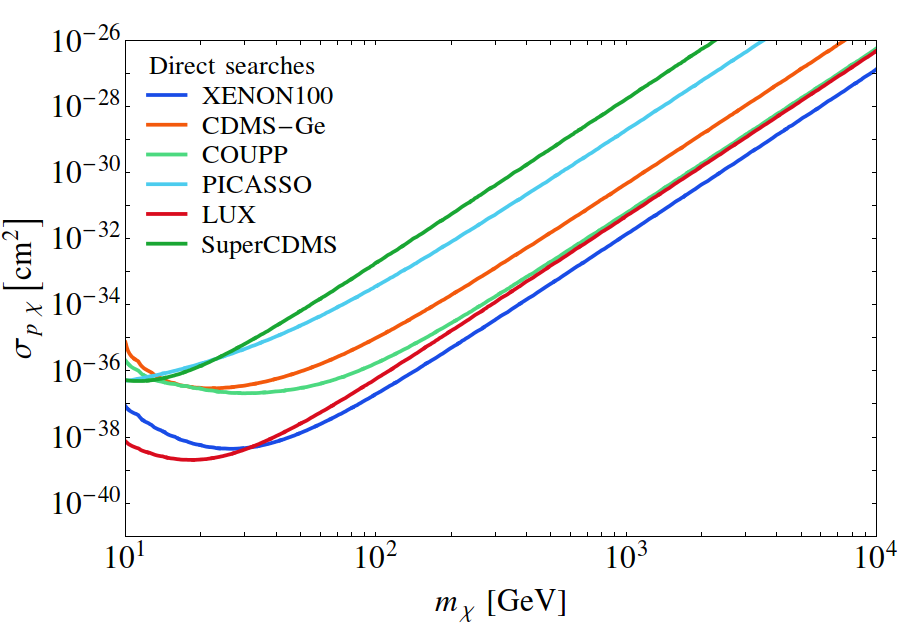} \hfill
  \includegraphics[width=0.49\textwidth]{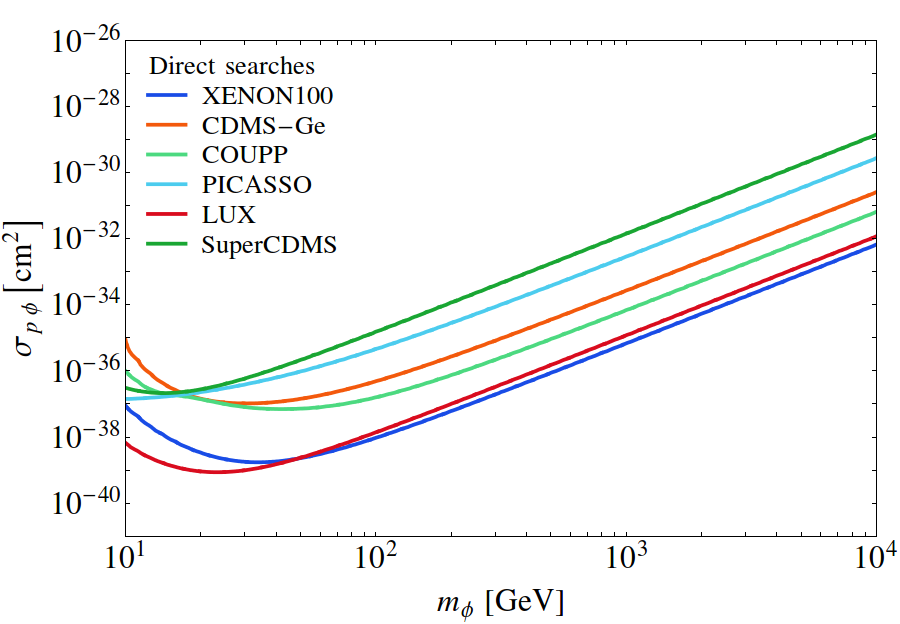}
  \caption{\new{Upper limits on the scattering cross section between proton and DM from direct detection experiments, for fermionic (\textit{left}) and scalar DM (\textit{right}).}}
  \label{fig:DD}
\end{figure}

\begin{figure}[h!]
\centering
\includegraphics[width=0.9\textwidth]{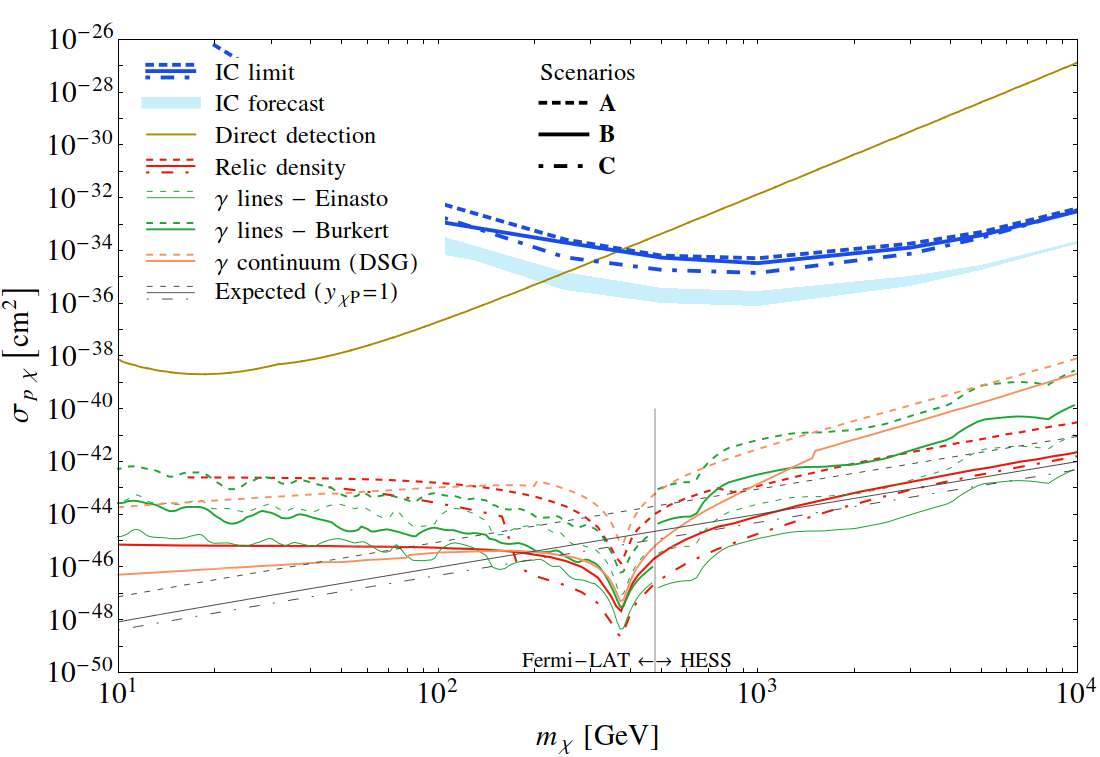}
\caption{
\new{
Upper bounds on $\spx$ from Direct Detection, IceCube and Indirect Detection experiments, for fermionic DM. 
The convolution of DD constraints is shown with a golden solid line. All the other constraints depend on the chosen scenario, which is identified by a specific dashing style. 
IC constraints are plotted with thick blue lines, and forecasts for a neutrino telescope with 300 times the exposure of IceCube are shown with a light blue band. 
With green and orange lines we show respectively the upper bounds from $\gamma$-ray lines and $\gamma$-ray continuum observations. To improve the readability, we use the same colour code for Fermi-LAT and HESS constraints on $\gamma$-ray lines which apply to different ranges of $\mx$, and we do not show the constraint for scenario C, being similar to the other two. 
The red lines show the prediction obtained by imposing that the relic density of $\chi$ and $\overline \chi$ equals the observed one.
We show also the expected signal for $\yxS=1$ (thin grey lines).
}
}
\label{fig:results fermion DM}
\end{figure}

The bounds from DD are obtained with the use of the software made available by the authors of Ref.~\cite{DelNobile:2013sia}, which allows to derive easily the bounds on a combination of non-relativistic operators for the interaction between DM and nucleons. 
\new{
We show in Fig.~\ref{fig:DD} the upper bounds on the scattering cross section between proton and DM for the six experiments included in the software of \cite{DelNobile:2013sia}, for the case of fermion and scalar DM. 
In the final comparison plots (Figs.~\ref{fig:results fermion DM}, \ref{fig:results scalar DM}) we show only the convolution of these exclusion limits, corresponding to a combination of the bounds from LUX and XENON-100.
}

Together with the current IC bounds obtained using the procedure explained above, we show a forecast for the bound that could be obtained by a similar neutrino telescope with 300 times the exposure. 
The sensitivity will scale with the square root of exposure \cite{IceCubeSearchMethodPaper}. 
While the exposure scales linearly with time and an improvement of $\sim300$ is unlikely for IceCube, future proposed neutrino telescopes such as KM3Net/ARCA \cite{Adrian-Martinez:2016fdl} may achieve a similar improvement in sensitivity faster, using larger volumes, better angular resolutions or improved analysis techniques. 
However, it is not clear if future neutrino detectors will target the sub TeV energy range in primary neutrinos that is crucial for Solar DM searches. 
% Together with the IC current bounds obtained with the procedure exposed above, we show a forecast for the bound that could be obtained by a neutrino telescope with an effective area 10 times bigger than the current IC one (for simplicity we take it constant with respect to $\mx$), an exposure time of 30 years (10 times more than the current exposure time of IC), and an improvement in sensitivity of a factor 3. 
% This leads to an improvement of IC bounds of a factor $\sqrt{10\cdot 10 \cdot 3}$. 

\new{
 We now consider the bounds from the observations of $\gamma$-rays in the sky. We recast the constraints from the searches for  spectral lines in the spectrum of $\gamma$-rays from the centre of the Milky Way from Fermi-LAT \cite{Ackermann:2015lka} and HESS \cite{Abramowski:2013ax}. 
As benchmark choices for the DM density profile we select the Einasto and Burkert profiles \cite{Baratella:2013fya}, which present a high and a low density in the centre of the Galaxy, respectively. 
The two experiments are sensitive to different energy ranges, thus constraining complementary intervals in $\mx$.  We show them with the same colour code in Figs.~\ref{fig:results fermion DM} and \ref{fig:results scalar DM}. 
We also examine the exclusion bounds from the the $\gamma$-ray continuum searches from a set of 15 Dwarf Spheroidal Galaxies (DSG) performed by Fermi-LAT \cite{Ackermann:2015zua}. 
A detailed description of our recasting procedure can be found in Appendix~\ref{sec: gamma rays}.
}

We also show, as a tentative reference point, the expected $\spx$ for a value of $\yxS =1$, from Eq.~\eqref{eq:xsec chi N to chi N}
and the  lines corresponding to the points of the parameter space where the relic energy density of the DM candidate through the freeze-out mechanism turns out to be $\Omega_{\rm DM} h^2 = 0.1196\pm0.0031$ \cite{Ade:2013zuv}.

Our results are summarized in Fig.~\ref{fig:results fermion DM}. 

%The conclusion we draw from Fig.~\ref{fig:results fermion DM} is that the double velocity suppression arising in the case of fermion DM and pseudoscalar mediator dramatically reduces  the experimental exclusion reach, and the constraints on the parameter of such a model are very weak. This outcome further motivates us to investigate the case of scalar DM.
The conclusion we draw from Fig.~\ref{fig:results fermion DM} is that the double velocity suppression arising in the case of fermion DM and pseudoscalar mediator dramatically reduces  the experimental exclusion reach \new{of DD and IC}. \new{Stronger constraints from $\gamma$-ray observations turn out to fall in the region favoured by the calculation of DM relic density. }
This outcome further motivates us to investigate the case of scalar DM.

\subsection{The case of scalar DM}
\begin{figure}[h!]
\centering
\includegraphics[width=0.9\textwidth]{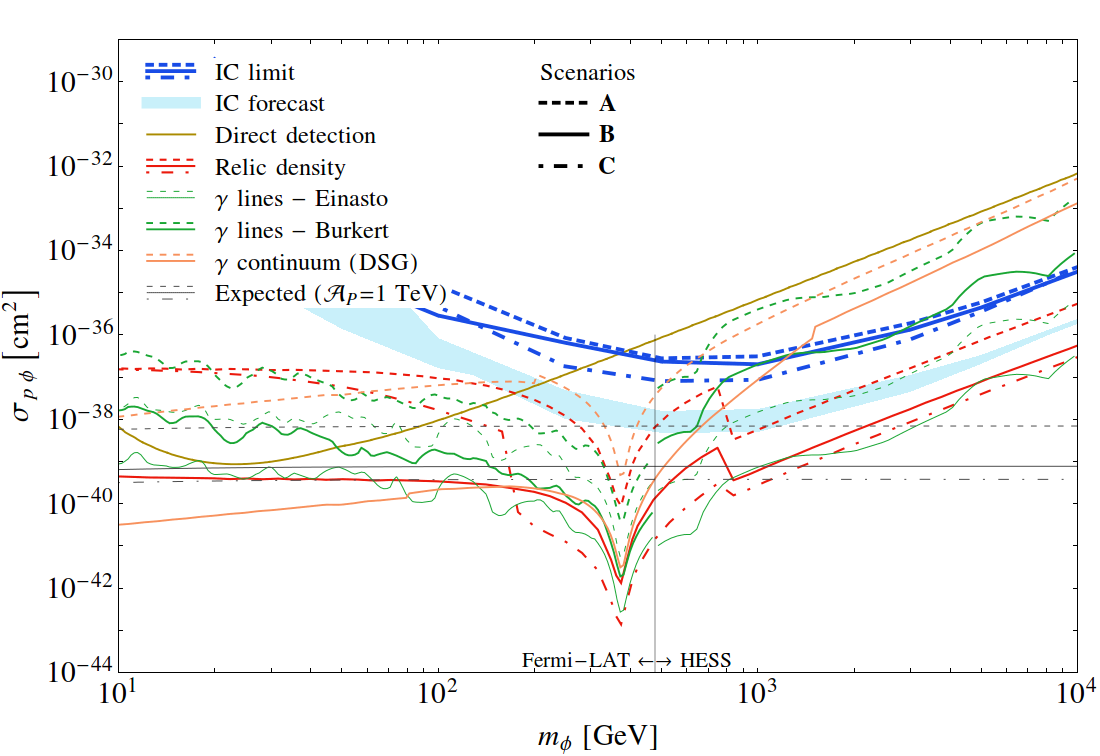}
\caption{The same as in figure \ref{fig:results fermion DM}, for the case of scalar DM.}
\label{fig:results scalar DM}
\end{figure}

In this case the non-relativistic operator for the interaction between the DM particle $\phi$ and nucleon is $\mathcal O_{10}^\text{NR}=i(\vec s_N \cdot \vec q)$, which is SD and suppressed by one power of DM velocity, instead of two as in the fermion DM case.
This behaviour alters significantly the experimental bounds, as apparent from Fig.~\ref{fig:results scalar DM}.

As a reference point, we show the predicted cross section for the  value $A_P=1$ TeV of the dimensionful coupling $A_P$ introduced in Eq.~\eqref{eq:L DM scalar}. The cross section $\spf$, reported in Eq.~(\ref{eq:xsec phi N to phi N}), is proportional to $A_P^2$. 
We conclude that, in the case in which the DM is a scalar particle, IC experimental bounds are important for two reasons: first, for masses
of the DM larger than ${\cal O}(300)$ GeV, IC constraints are stronger than the bounds coming from DD experiments and, secondly, 
they bound $A_P$ to be not larger than a few TeV. Forecasts for a possible future neutrino telescope lower this bound to around 1 TeV in the region $(m_\phi \sim 0.1\div 1)$ TeV, improving significantly the bounds from DD experiments in the mass region close to $\mS$.
\new{
The reach of the exclusion limits from the observation of $\gamma$-ray lines is  affected by astrophysical uncertainties, mainly the DM density profile. 
In the region $\mf\gtrsim 500$ GeV, HESS yields are comparable or weaker to IC for DM profiles such as Burkert or Isothermal, and stronger for steeper profiles as Einasto. 
Upper limits from the observations of DSG, on the other hand, fall above IC ones.
}

\section{Conclusions}
\label{sec: conclusions}
Should future  data collected by the LHC collaborations ATLAS and CMS  confirm the  existence of a new resonant state  with a mass of around 750 GeV, the era of physics beyond the SM would start. 
Following this wishful route, one can imagine that the resonance acts a portal into the DM sector. 
By identifying three benchmark models, in this paper we have investigated the possibility that the resonance is caused by a pseudoscalar particle which also couples to either fermionic or scalar DM. 
Motivated by the fact that  the bounds from solar DM searches are generally more stringent than DD experiments when  the DM-nucleon scattering depends on the spin of the nucleus, we have analyzed the
bounds coming from the search for neutrinos originated from DM annihilations in the Sun performed by the IC collaboration\new{, and we have compared them with Direct Detection experiments and $\gamma$-ray observations}. 
Our findings indicate that the IC data  provide constraints stronger than the DD experiments for DM masses larger than a few hundred GeV. Furthermore,  if the DM is a scalar particle, the IC data limit the cross section between the DM and the proton to values close to the predicted ones for natural values of the parameters.

%
%%% ACKNOWLEDGMENTS %%%
\acknowledgments
 D.R.\ and A.R.\ are  supported by the Swiss National
Science Foundation (SNSF), project ``Investigating the Nature of Dark Matter'' (project number: 200020\textunderscore{}159223).

%
%
%
%

%%% APPENDICES %%%
%
\appendix

\section{Relevant formul\ae\ for scattering with protons, annihilation of DM and computation of relic density}
\label{sec:appendix A}
We report in this section the relevant formul\ae\ for the computation of cross sections for DD, for DM annihilations (which are relevant for IC), and of the relic density of DM.

\subsection{Elastic scattering between proton and DM}

\paragraph{Fermionic DM}
The effective operator for the interaction between $\chi$ and a nucleon $N$ is (throughout this section we follow the notation of Ref. \cite{DelNobile:2013sia})
\begin{equation}
\label{eq:O chi N}
\lambda_\chi {\cal O}_{4}^{\chi  N} =\frac{\yxS\, c_N}{\mS^2} (\overline \chi \gamma_5 \chi) \, (\overline N \gamma_5 N) \,,
\end{equation}
where the coefficient $\yxS$ is introduced in Eq.~\eqref{eq:L DM fermion}, and we define $c_N$ as
\begin{equation}
\label{eq:cN}
c_N= \sum_{q=u,d,s} \frac{m_N}{m_q} \Delta_q^{(N)} 
          \left[ \yqS -\frac{\cgg}{\Lambda} \overline m
          + \sum_{q'=u,d,s} y_{q' \S} \frac{\overline m}{m_{q'}}    \right] \,,
\end{equation}
where $\overline m = \big( \tfrac{1}{m_u}+\tfrac{1}{m_d}+\tfrac{1}{m_s} \big)^{-1}$, and $\Delta_q^{(N)}$ are the nucleon spin form factors for the quark $q$ (see Ref.  \cite{DelNobile:2013sia} for a compilation of their numerical values according to various references).

The cross section for the elastic scattering $\chi N \to \chi N$ in the low velocity limit turns out to be (the same result holds for the process $\overline \chi N \to \overline \chi N$)
\begin{equation}
\label{eq:xsec chi N to chi N}
\sigma_{N \chi} = \frac{1}{3\pi} \lambda_\chi^2\, \mu_N^2 \frac{\mx^2}{m_N^2}\, \vDM^4 +\mathcal O \big(\vDM^6 \big) \,,
\end{equation}
where $\vDM$ is the DM particle velocity in the centre-of-mass frame (which we assume to be 220 km/s), and $\mu_N=\tfrac{\mx m_N}{\mx+m_N}$ is the DM-nucleon reduced mass.

We notice that this cross section is suppressed by the fourth power of the velocity of DM, as a result of the presence of the two pseudoscalar Lorentz bilinears in Eq.~\eqref{eq:O chi N} which in the non-relativistic limit reduce each to the scalar product of the spin of the correspondent particle and of $\vec q$.  

\paragraph{Scalar DM} 
The effective operator for the interaction between the DM particle $\phi$ and a nucleon is
\begin{equation}
\label{eq:O phi N}
\lambda_\phi {\cal O}_{2}^{\phi N} =\frac{A_P\, c_N}{\mS^2} (\phi^* \phi) \, (\overline N i \gamma_5 N) \,,
\end{equation}
where $A_P$ is the parity-odd dimensionful coefficient introduced in Eq.~\eqref{eq:L DM scalar}, and $c_N$ is defined in Eq.~\eqref{eq:cN}.
The corresponding cross section in the low velocity limit is
\begin{equation}
\label{eq:xsec phi N to phi N}
\sigma_{N \phi} = \frac{1}{8\pi} \lambda_\phi^2\, \mu_N^2 \frac{1}{m_N^2}\, \vDM^2 +\mathcal O\big(\vDM^4 \big) \,.
\end{equation}
We point out that in this case $\sigma_{N\phi}$ is suppressed just by the second power of $\vDM$, accordingly to the fact that ${\cal O}_{2}^{\phi N}$ contains only one pseudoscalar Lorentz bilinear.

\subsection{Annihilation of DM pairs into SM channels}
The squared matrix elements, summed over the polarisations of the final states, for the two-body decays of $P$  are the following \cite{D'Eramo:2016mgv} (we denote by $\tW$ the Weinberg angle):
\begin{align}
\label{eq:M^2 first}
\AMS{gg} & = 64 \frac{\cgg^2}{\Lambda^2} \, s^2 \,, \\
\AMS{ZZ} & = \frac{8\sin^4\tW\, \cBB^2}{\Lambda^2} \, s^2 \left( 1- \frac{4m_Z^2}{s}\right) \,, \\
\AMS{Z\gamma} & = \frac{4 \cos^2\tW \, \sin^2\tW\, \cBB^2}{\Lambda^2} \, s^2 \left( 1- \frac{m_Z^2}{s}\right)^2  \,, \\
\AMS{\gamma\gamma} & = \frac{8 \cos^4\tW \, \cBB^2}{\Lambda^2} \, s^2 \\
\AMS{\overline q q} & = 6\, \yqS ^2\, s  \,, \\
\AMS{\overline \chi \chi} & = 2\, \yqS ^2\, s \label{eq:M^2 P to chi chi} \,,\\
\AMS{\overline \phi\phi} & = A_P^2 \label{eq:M^2 P to phi phi} \,.
\end{align}
The partial width of $\S$ into a final state composed of two particles $ij$ is then given by
\begin{equation}
\label{eq:width P to ij}
\Gamma_{\S \to ij} = s_{ij} \,\frac{\AMS{ij}}{16\pi\,\mS}
    \sqrt{1-2\frac{m_i^2+m_j^2}{s} +\frac{(m_i^2-m_j^2)^2}{s^2}}\,,
\end{equation}
where $s_{ij}$ is a symmetry factor equal to $1/2$ if $i$ and $j$ are identical particles, or $1$ otherwise. 

We can then write the cross section for the annihilation of $\overline \chi \chi $ into a final state $ij$ at a centre-of-mass energy $s$ as
\begin{equation}
\label{eq:xsec chi chi to i j}
\sigma_{\overline \chi \chi \to ij} = s_{ij} \, \simplify{\frac{\yxS^2}{32\pi}}
  \frac{\AMS{ij}}{ \simplify{(s-\mS^2)^2+\GS^2\mS^2} }
    \frac{\sqrt{1-2\frac{m_i^2+m_j^2}{s} +\frac{(m_i^2-m_j^2)^2}{s^2}}}{ \simplify{\sqrt{1-\frac{4\mx^2}{s}}} } \,.
\end{equation}
We highlight with a shade of grey the terms that are common to all final states, and simplify when computing the branching ratios for the annihilation of DM pairs into SM channels. 
These are indeed the relevant quantities for computing the neutrino fluxes for IC. 
This interesting simplification makes IC bound independent of the coupling of $\S$ to DM, of $\mS$ and of $\GS$.

The analogue formula to Eq.~\eqref{eq:xsec chi chi to i j} for the case of scalar DM is:
\begin{equation}
\label{eq:xsec phi phi to i j}
\sigma_{\overline \phi \phi \to ij} = s_{ij} \, \simplify{\frac{1}{64\pi} \frac{A_P^2}{s}}
  \frac{\AMS{ij}}{ \simplify{(s-\mS^2)^2+\GS^2\mS^2} }
    \frac{\sqrt{1-2\frac{m_i^2+m_j^2}{s} +\frac{(m_i^2-m_j^2)^2}{s^2}}}{ \simplify{\sqrt{1-\frac{4\mf^2}{s}}} } \,.
\end{equation}
The centre-of-mass energy $s$, when applying Eq.s~\eqref{eq:xsec chi N to chi N} and \eqref{eq:xsec phi N to phi N} to DM annihilations inside the Sun, has to be evaluated with the typical kinetic energy of DM particles in the Sun. 
We notice that all the squared matrix elements in Eqs.~\eqref{eq:M^2 first}-\eqref{eq:M^2 P to phi phi} in the low velocity expansion have a non vanishing constant term.
We assume as a reference kinetic energy the thermal one inside the core of the Sun, around $\sim 1$ keV (corresponding to a velocity $10^{-4}$ for $\mx\sim 100$ GeV).

\subsection{Relic density of DM via freeze-out}
\label{sec:relic}
We collect the main formul\ae\ needed to compute the actual relic abundance of DM. For a thorough discussion, we refer to  Refs. \cite{Gondolo:1990dk, Kolb:1990vq}.

By defining $x=\mDM/T$, the expression for the thermally averaged cross section $\asv$ at a temperature $T$ reads \cite{Gondolo:1990dk}
\begin{equation}
\label{eq:asv}
\asv = \frac{x}{8\,\mDM^5}\frac{1}{\big(\bes{2}(x)\big)^2} \int_{4\mDM^2}^\infty \sigma_\text{ann} \sqrt s \, \left(s-4 \mDM^2 \right) \,\bes{1}\left(\frac{x\sqrt s}{\mDM} \right)\, \d s\,,
\end{equation}
where $\bes{i}$ is the modified Bessel function of order $i$.

The relic abundance is then obtained from \cite{Kolb:1990vq,Busoni:2014gta}
\begin{equation}
\label{eq:Omega DM}
\ODM h^2 = \Omega_\chi h^2 + \Omega_{\bar\chi}h^2= \frac{2\times 1.04\times 10^9 \, {\rm GeV}^{-1} \mDM}{M_{\rm Pl} \int_{T_0}^{T_f} g_\star^{1/2} \asv {\rm d}T},
\end{equation}
where $T_f$ is the freeze-out temperature, $T_0$ is the present temperature, $g_\star$ is the degrees of freedom parameter as a function of the temperature and the factor of 2 accounts for the fact that the total DM density is the sum of the density of DM particles and antiparticles.
Note that here we approximate $g_\star = g_{\rm eff} = h_{\rm eff}$, where $g_{\rm eff}$ and $h_{\rm eff}$ are the number of degrees of freedom that enters the definition of the energy density and of the entropy density respectively \cite{Gondolo:1990dk}.

The freeze-out temperature $T_f$ (or, equivalently, $x_f$) is defined to be the temperature at which the quantity $Y=n/s$ differs from its equilibrium value by $Y-Y_{\rm eq} = c \, Y_{\rm eq}$.
With a standard notation we indicate by $n$ and $s$ the number density of DM particles and the entropy density of the Universe.
The value of $x_f$ is obtained by
\begin{equation}
\label{eq:xf}
e^{x_f} = \frac{\sqrt{\frac{45}{8}} g \mDM M_{\rm Pl} c(c+2) \asv}{2 \pi^3 g_\star^{1/2} \sqrt{x_f}},
\end{equation}
where $g = 1 \textrm{ (scalar)}, 2 \textrm{ (spinor)}$ accounts for the number of spin states of the DM particle. Following Ref.~\cite{Gondolo:1990dk} we take $c=1.5$.

The thermally averaged cross section appearing in the denominator of Eq.~\eqref{eq:Omega DM} can be expanded as a function of $x$. This approximation can be slightly inaccurate when a resonance is excited \cite{Griest:1990kh}, because for $\mDM$ just below the resonance also higher orders of the expansion of $\sigma_\text{ann}(v)$ at low velocity can matter. For this reason, we perform our computation without expanding Eq.~\eqref{eq:asv} in powers of $\vDM$.

We report the results for the expansion of $\svx{ij}$ in series of $\vMol$, to show that the term of order zero in the velocity is always non vanishing. Therefore, the final value for the cross section does not depend strongly on the numerical value assumed for DM velocity.

For the case of fermionic DM, the expanded cross sections read
\begin{small}
\begin{multline}
\svx{gg} =
  \frac{
    32 \yxS^2 \cgg^2  \mx^4 
  }{
    \pi \Lambda^2 \big[ (\mS^2- 4 \mx^2)^2 + \GS^2 \mS^2 \big]
  }
  + \vMol^2 \frac{
    16 \yxS^2 \cgg^2 \mS^2 \mx^4 ( \mS^2 -4 \mx^2 + \GS^2 )
  }{
    \pi \Lambda^2 \big[ (\mS^2- 4 \mx^2)^2 + \GS^2 \mS^2 \big]^2
  }\, ,
\end{multline}
\vspace{-2em}
\begin{multline}
\label{eq:XX to gamma gamma}
\svx{\gamma \gamma} =
  \frac{
    4 \yxS^2 c_{\gamma\gamma}^2  \mx^4 
  }{
    \pi \Lambda^2 \big[ (\mS^2- 4 \mx^2)^2 + \GS^2 \mS^2 \big]
  }
  + \vMol^2  \frac{
    2 \yxS^2 c_{\gamma\gamma}^2  \mS^2 \mx^4 ( \mS^2 -4 \mx^2 + \GS^2 )
  }{
    \pi \Lambda^2\big[ (\mS^2- 4 \mx^2)^2 + \GS^2 \mS^2 \big]^2
  }\, ,
\end{multline}
\vspace{-2em}
\begin{multline}
\label{eq:XX to Z gamma}
\svx{Z \gamma} =
  \frac{
    2 \yxS^2 c_{Z\gamma}^2  \mx^4 
  }{
    \pi \Lambda^2 \big[ (\mS^2- 4 \mx^2)^2 + \GS^2 \mS^2 \big]
  } \left(1-\frac{m_Z^2}{4 \mx^2} \right)^3 
  + \vMol^2 \left(1-\frac{m_Z^2}{4 \mx^2} \right)^2  \\
  \cdot \frac{
    \yxS^2 c_{Z\gamma}^2  \mx^2}
    {  8 \pi \Lambda^2 }
  \left[ \frac{
   8 \mx^2 + m_Z^2}{
    (\mS^2- 4 \mx^2)^2 + \GS^2 \mS^2 
  }
  + \frac{8 \mx^2 (\mS^2-4 \mx^2) (4 \mx^2-m_Z^2)
    }{
    \big[ (\mS^2- 4 \mx^2)^2 + \GS^2 \mS^2 \big]^2
  } \right]\, ,
\end{multline}
\vspace{-2em}
\begin{multline}
\svx{Z Z} =
  \frac{
    4 \yxS^2 c_{ZZ}^2  \mx^4 
  }{
    \pi \Lambda^2  \big[ (\mS^2- 4 \mx^2)^2 + \GS^2 \mS^2 \big]
  } \left(1-\frac{m_Z^2}{\mx^2} \right)^{3/2}
  + \vMol^2 \left(1-\frac{m_Z^2}{\mx^2} \right)^{1/2}  \\
  \cdot \frac{
    \yxS^2 c_{ZZ}^2  \mx^2}{
    2 \pi \Lambda^2
    } \left[ \frac{
 4 \mx^2 - m_Z^2 }{
    (\mS^2- 4 \mx^2)^2 + \GS^2 \mS^2 
  }
  + \frac{ 
  16 \mx^2 (\mS^2 - 4 \mx^2) (\mx^2 - m_Z^2)
  }{
  \big[ (\mS^2- 4 \mx^2)^2 + \GS^2 \mS^2 \big]^2
  } \right]\, , 
\end{multline}
\vspace{-2em}
\begin{multline}
\svx{\overline q q} =
  \frac{
    3 \yxS^2 \yqS^2 \mx^2
  }{
    2 \pi \big[ (\mS^2- 4 \mx^2)^2 + \GS^2 \mS^2 \big]  
  } \left(1-\frac{m_q^2}{\mx^2} \right)^{1/2} 
  +  \vMol^2  \left(1-\frac{m_Z^2}{\mx^2} \right)^{-1/2}
   \\
  \cdot
  \frac{
    3 \yxS^2 \yqS^2 \mx^2}
    {16 \pi }
  \left[
  \frac{
    m_q^2 - 2 \mx^2
  }{
    (\mS^2- 4 \mx^2)^2 + \GS^2 \mS^2 
  }
  +\frac{
    16 \mx^2 (\mS^2-4 \mx^2) (m_q^2-\mx^2) 
  }{
    \big[ (\mS^2- 4 \mx^2)^2 + \GS^2 \mS^2 \big]^2
  } 
  \right]\, .
\end{multline}
\end{small}
%
%\end{document}

In the case of scalar DM, the results are
%\documentclass[11pt,a4paper]{article}
%\usepackage[utf8x]{inputenc}
%\usepackage{ucs}
%\usepackage[english]{babel}
%\usepackage{amsmath}
%\usepackage{amsfonts}
%\usepackage{amssymb}
%\usepackage[left=2.5cm,right=2.5cm,top=3cm,bottom=3cm]{geometry}
%
%% Commands for notation
%%
%\renewcommand{\d}{\mathrm d}
%% LHC and couplings
%\renewcommand{\S}{P} % in case we want to label it as P
%\newcommand{\mS}{m_\S}
%\newcommand{\GS}{\Gamma_\S}
%\newcommand{\cgg}{c_{gg}}
%\newcommand{\cBB}{c_{BB}}
%\newcommand{\yxS}{y_{\chi\S}}
%\newcommand{\yqS}{y_{q\S}}
%\newcommand{\yuS}{y_{u\S}}
%\newcommand{\ydS}{y_{d\S}}
%\newcommand{\ysS}{y_{s\S}}
%\newcommand{\ybS}{y_{b\S}}
%\newcommand{\ytS}{y_{t\S}}
%\newcommand{\mx}{m_\chi}
%\newcommand{\mf}{m_\phi}
%\newcommand{\tW}{\theta_W}
%% DM
%\newcommand{\vMol}{v_\text{M\o l}}
%\newcommand{\svx}[1]{\sigma(\overline \chi \chi\to #1) \cdot \vMol}
%\newcommand{\svf}[1]{\sigma(\overline \phi \phi\to #1) \cdot \vMol}
%
%
%\begin{document}
\begin{small}
\begin{multline}
\svf{gg} =
  \frac{
    4 A_P^2 \cgg^2  \mf^2 
  }{
    \pi \Lambda^2 \big[ (\mS^2- 4 \mf^2)^2 + \GS^2 \mS^2 \big]
  }
  + \vMol^2 \frac{
    A_P^2 \cgg^2 \mf^2 ( \mS^4 -16 \mf^4 + \GS^2 \mS^2 )
  }{
    \pi \Lambda^2 \big[ (\mS^2- 4 \mf^2)^2 + \GS^2 \mS^2 \big]^2
  }\, ,
\end{multline}
\vspace{-2em}
\begin{multline}
\svf{\gamma \gamma} =
  \frac{
    A_P^2 c_{\gamma\gamma}^2  \mf^2
  }{
    2\pi \Lambda^2 \big[ (\mS^2- 4 \mf^2)^2 + \GS^2 \mS^2 \big]
  }
  + \vMol^2  \frac{
    A_P^2 c_{\gamma\gamma}^2  \mf^2 ( \mS^4 -16 \mf^4 + \GS^2\mS^2  )
  }{
    8 \pi \Lambda^2\big[ (\mS^2- 4 \mf^2)^2 + \GS^2 \mS^2 \big]^2
  }\, ,
\end{multline}
\vspace{-2em}
\begin{multline}
\svf{Z \gamma} =
  \frac{
    A_P^2 c_{Z\gamma}^2  \mf^2 
  }{
    4 \pi \Lambda^2 \big[ (\mS^2- 4 \mf^2)^2 + \GS^2 \mS^2 \big]
  } \left(1-\frac{m_Z^2}{4 \mf^2} \right)^3 
  + \vMol^2 \left(1-\frac{m_Z^2}{4 \mf^2} \right)^2  \\
  \cdot \frac{
    A_P^2 c_{Z\gamma}^2 }
    {  32 \pi \Lambda^2 }
  \left[ \frac{
   2 \mf^2 + m_Z^2}{
    (\mS^2- 4 \mf^2)^2 + \GS^2 \mS^2 
  }
  +\frac{4 \mf^2 (\mS^2-4 \mf^2) (4 \mf^2-m_Z^2)
    }{
    \big[ (\mS^2- 4 \mf^2)^2 + \GS^2 \mS^2 \big]^2
  } \right]\, ,
\end{multline}
\vspace{-2em}
\begin{multline}
\svf{Z Z} =
  \frac{
    4 A_P^2 c_{ZZ}^2  \mf^2 
  }{
    \pi \Lambda^2  \big[ (\mS^2- 4 \mf^2)^2 + \GS^2 \mS^2 \big]
  } \left(1-\frac{m_Z^2}{\mf^2} \right)^{3/2}
  + \vMol^2 \left(1-\frac{m_Z^2}{\mf^2} \right)^{1/2}  \\
  \cdot \frac{
    A_P^2 c_{ZZ}^2 }{
    16 \pi \Lambda^2
    } \left[ \frac{
 2 \mf^2 + m_Z^2 }{
    (\mS^2- 4 \mf^2)^2 + \GS^2 \mS^2 
  }
  + \frac{ 
  16 \mf^2 (\mS^2 - 4 \mf^2) (\mf^2 - m_Z^2)
  }{
  \big[ (\mS^2- 4 \mf^2)^2 + \GS^2 \mS^2 \big]^2
  } \right]\, ,
\end{multline}
\vspace{-2em}
\begin{multline}
\svf{\overline q q} =
  \frac{
    3 A_P^2 \yqS^2
  }{
    16 \pi \big[ (\mS^2- 4 \mf^2)^2 + \GS^2 \mS^2 \big]  
  } \left(1-\frac{m_q^2}{\mf^2} \right)^{1/2} 
  +  \vMol^2  \left(1-\frac{m_Z^2}{\mf^2} \right)^{-1/2}
   \\
  \cdot
  \frac{
    3 A_P^2 \yqS^2 }
    {128 \pi \mf^2 }
  \left[
  \frac{
    m_q^2
  }{
    (\mS^2- 4 \mf^2)^2 + \GS^2 \mS^2 
  }
  -\frac{
    16 \mf^2 (\mS^2-4 \mf^2) (m_q^2-\mf^2) 
  }{
    \big[ (\mS^2- 4 \mf^2)^2 + \GS^2 \mS^2 \big]^2
  } 
  \right]\, .
\end{multline}
\end{small}
%
%\end{document}

\subsection{Constraints from observations of $\gamma$-rays}
\label{sec: gamma rays}
\subsubsection*{Search for $\gamma$-ray spectral lines from the centre of the Milky Way}
In this section we provide further details about the procedure we adopted in order to recast the exclusion bounds on DM annihilations coming from the search for $\gamma$-ray spectral lines from the centre of the Milky Way.

The annihilation of two DM particles\footnote{For simplicity of notation, we will consider throughout this appendix a Dirac fermion $\chi$ as DM particle.} $\chi$ and $\overline \chi$ into $\gamma\gamma$, in the limit $\vDM\to 0$, provides two photons of energy $\Eg = \mx$. The process $\chi \overline \chi \to Z\gamma$ gives only one photon, with an energy $\Eg = \mx \left(1- \tfrac{m_Z^2}{4\mx^2} \right)$. The constraints provided by Fermi and HESS on $\asv$ however are derived from searches for monochromatic spectral lines. Nevertheless for $\mx$ larger than 100 GeV, the relative separation in the energies of $\gamma$s from the two processes is smaller than the energy resolution of the instrument and consequently, constraints can be derived using the following method. For $\mx$ smaller than 100 GeV, the second term in the LHS of Eq.~\eqref{eq:gammarescale} is ignored to obtain a conservative bound.

The fluxes of photons obtained in our model for a DM mass $\mx$, to be compared with the exclusion limit $\asv_{\gamma \gamma }^\text{limit}$ provided by Fermi-LAT or HESS which assume that only $\chi \overline \chi \to \gamma\gamma$ occurs, is given by
\begin{equation}
\label{eq:gammarescale}
\asv_{\gamma \gamma } (\mx )+\frac 12 \cdot \asv_{Z\gamma }\left(\frac{1}{2} \left(\mx+\sqrt{\mx^2+m_Z^2}\right)\right)
\leq 
\asv_{\gamma \gamma }^\text{limit} (\mx) \,,
\end{equation}
where the arguments of $\asv$ inside brackets refer to the DM mass to be inserted into Eqs.~\eqref{eq:XX to gamma gamma}, \eqref{eq:XX to Z gamma}.

The Fermi-LAT analysis, performed with the data collected over 6 years by the Large Area Telescope hosted by the satellite Fermi \cite{Ackermann:2015lka}, identifies different signal regions depending on the DM density profile under consideration. 
They select the region R16 (a cone with an opening angle $\theta=16\deg$ around the centre of the Milky Way) for the profile Einasto, and the region R90 (corresponding to $\theta=90\deg$, i.~e.~half of the sky around the centre) for the profile Isothermal. 
We want to recast the limits assuming the DM profile Burkert, characterised by a lower DM density in the centre of the galaxy. 
Thus we compute the integral of the $J$-factors in the region $\theta \leq 90\deg $ with the tables provided by \cite{Cirelli:2010xx} (to which we refer for the definitions of the DM density profiles and of the $J$-factors), and we use it to rescale the bound computed by Fermi-LAT assuming the Isothermal profile.

The sensitivity of LAT to $\gamma$-rays of energies up to $\sim 300$ GeV is complemented by the sensitivity of the telescope HESS, located in Namibia, which provides important bounds on the annihilation of DM into $\gamma$ rays for $\mx\gtrsim 500$ GeV \cite{Abramowski:2013ax}. HESS observes a cone of $1\deg$ around the centre of the Milky Way, and recasts the exclusion limits assuming the Einasto profile. Analogously to what we did in the previous case, in order to get the corresponding limits with a Burkert profile we rescale them by computing the ratio of the integrated $J$-factors.

\subsubsection*{Observation of the $\gamma$-ray continuum from Dwarf Spheroidal Galaxies}
Fermi-LAT performed \cite{Ackermann:2015zua} an analysis of the $\gamma$-ray spectrum coming from 15 DSGs. These particular galaxies are characterised by an higher density of DM particles than ordinary galaxies, thus they represent an ideal target to look for secondary $\gamma$-rays resulting from the primary products of DM annihilations.

A careful recast of the analysis of \cite{Ackermann:2015zua} would require the computation of the spectrum of $\gamma$-rays from DM annihilations (which can be done through the tables provided by \cite{Cirelli:2010xx}), and the knowledge of the exclusion limits on the fluxes of $\gamma$-rays. 
This information is not available, but we can make a good approximation in our scenarios by identifying, in suitable intervals of $\mx$, the leading annihilation channel providing secondary photons (for example, in scenario B we consider only the annihilation channels $\overline \chi\chi \to \overline b b$ for $\mx<1.5$ TeV and $\overline \chi\chi \to Z\gamma$ for $\mx>1.5$ TeV, see Fig.~\ref{fig:BR B}). 
The exclusion limits are recast by \cite{Ackermann:2015zua} as limits on the thermally averaged annihilation cross section into some specific channels, assuming in every case that they are the only annihilation channels of DM. 
These include $u\overline u$, which provides basically the same spectrum as $gg$, and $W^+W^-$, which yields analogous fluxes as $ZZ$, $t\overline t$, $b\overline b$, and the double of the flux of $Z\gamma$ \cite{Cirelli:2010xx}. 
These considerations allow us to perform the recast of Fermi-LAT observations by imposing that our main annihilation channel (for a given scenario and range of $\mx$) equates the exclusion limit of \cite{Ackermann:2015zua}.

%
%%% BIBLIOGRAPHY with BIBTEX %%%
%
%\nocite*{} % to print all the entries in the bib file, even if they are not cited in the text
\bibliographystyle{JHEP}
\bibliography{biblio_Res-IC}

%
%%% BIBLIOGRAPHY with the usual latex command %%%
%
%\begin{thebibliography}{999}
%\end{thebibliography}

\end{document}